\documentclass[12pt]{article}
\usepackage{amsmath,amssymb}
\usepackage[utf8]{inputenc}
\usepackage{graphicx}
\usepackage[margin=2cm]{geometry}

\renewcommand{\vec}{\boldsymbol}

\newcommand{\cald}{\,\mathrm{d}}

\DeclareMathOperator{\sinc}{sinc}

\begin{document}

\title{Closed form expressions for gravitational multipole moments of elementary solids}
\author{\small Julian Stirling$^{1,}$\footnote{Email: {\tt j.stirling@bath.ac.uk}} \hspace{1ex}and Stephan Schlamminger$^2$\\
\small $^1$ Department of Physics, University of Bath, Claverton Down, Bath, BA2 7AY, UK\\
\small $^2$ National Institute of Standards and Technology, 100 Bureau Drive, Gaithersburg, MD 20899, USA }
\maketitle

%\ead{}

%\noindent{\it Keywords}: Newtonian gravitation; multipole expansion; spherical harmonics; solid harmonics

\begin{abstract}
Perhaps the most powerful method for deriving the Newtonian gravitational interaction between two masses is the multipole expansion. Once inner multipoles are calculated for a particular shape this shape can be rotated, translated, and even converted to an outer multipole with well established methods. The most difficult stage of the multipole expansion is generating the initial inner multipole moments without resorting to three dimensional numerical integration of complex functions. Previous work has produced expressions for the low degree inner multipoles for certain elementary solids. This work goes further by presenting closed form expressions for all degrees and orders. A combination of these solids, combined with the aforementioned multipole transformations, can be used to model the complex structures often used in precision gravitation experiments.
\end{abstract}

\section{Introduction}
In the field of precision gravitational measurements, the measurements and its associated analysis are often only half of the battle in producing a result. The other half comes from computing the theoretical Newtonian gravitational interaction for comparison. Computation of gravitational fields, forces, and torques can be accomplished by calculating sextuple integrals over the volumes of mass pairs, and summing for all pairs of source and test masses. Even with advanced methods to reduce these sextuple integrals to quadruple integrals  \cite{ChenCookBook,StirlingNJP2017}, for certain elementary solids, this is extremely computationally intensive, especially considering that for many measurements this needs to be entirely recalculated for multiple source mass positions. More efficient methods are available for systems with favourable symmetries \cite{Cohl1999,Selvaggi2008}.

An elegant method to compute gravitational interactions is to expand the problem in terms of regular solid harmonics ${r_\mathrm{i}}^l Y_{lm}(\theta_\mathrm{i},\phi_\mathrm{i})$, or more precisely its complex conjugate ${r_\mathrm{i}}^l Y^\ast_{lm}(\theta_\mathrm{i},\phi_\mathrm{i})$, of the masses closest to the origin of the chosen coordinate system and the irregular solid harmonics  ${r_\mathrm{o}}^{-(l+1)} Y_{lm}(\theta_\mathrm{o},\phi_\mathrm{o})$, of the masses furthest from this origin. Where $\vec{r_\mathrm{i}} := (r_\mathrm{i},\theta_\mathrm{i},\phi_\mathrm{i})$, and $\vec{r_\mathrm{o}} := (r_\mathrm{o},\theta_\mathrm{o},\phi_\mathrm{o})$ are vectors to positions inside the inner and outer masses respectively. Triple integrals of $\rho(\vec{r_\mathrm{i}}){r_\mathrm{i}}^l Y^\ast_{lm}(\theta_\mathrm{i},\phi_\mathrm{i})$ over the volumes of the inner masses are referred to as the inner multipoles $q_{lm}$, where $\rho(\vec{r})$ is the mass density. Whereas triple integrals of $\rho(\vec{r_\mathrm{o}}){r_\mathrm{o}}^{-(l+1)} Y_{lm}(\theta_\mathrm{o},\phi_\mathrm{o})$ over the volumes of the outer masses are referred to as the outer multipoles $Q_{lm}$. The convergence condition for this expansion is that  $r_\mathrm{i} < r_\mathrm{o}$ for all positions integrated over. The gravitational potential energy of the system can be calculated as
\begin{equation}
 V = -4\pi G \sum_{l=0}^\infty\sum_{m=-l}^l \frac{1}{2l+1} q_{lm}Q_{lm}\,.
\end{equation}

At first glance, an infinite sum over pairs of triple integrals is not necessarily a significant advance over a brute force calculation of the sextuple integrals. The power of the multipole expansion becomes apparent when considering complex experiments with multiple source and test masses. The multipole moments can be calculated for each individual mass just once and then used with all masses it interacts with, as other masses change no recalculation is needed. Also multipole moments can easily undergo translations \cite{DUrso1997} and rotations \cite{Jackson,QuantTheoryAngMom}. As such, when calculating the effect of a mass moving, very few new calculations are needed. Furthermore, outer multipoles can be computed from inner multipoles of the same shape \cite{Trenkel1999}. Utilising the multipole transformations, the only other calculations needed are the inner multipole moment of each mass at an arbitrary location, which is easy to calculate.  Forces\cite{StirlingPRD2017} and torques\cite{Newman2014} can also be directly calculated from these multipoles.

Efficient calculation of inner multipole moments is, as such, of great value. Low-degree ($l\leq5$) inner multipole moments have been calculated individually for each order ($m$) for a number of elementary solids \cite{Adelberger2006}. For higher degrees, however, either numerical methods must be employed or each order must be calculated explicitly. In this work we develop closed form solutions for the inner multipole moments a number of solids. These, combined with the multipole transformations, can be used to calculate gravitational interactions between complex apparatus to any required accuracy with relative ease.

\section{Closed forms expressions for inner multipoles of homogeneous solids}\label{Sec:ClosedForm}

For calculating inner multipoles is is helpful to write the regular solid harmonics in the cylindrical coordinate system. From Eqn.~4.28 in Ref  \cite{ThompsonAngMom}, the solid harmonics are given in Cartesian coordinates. It is trivial to convert this form into cylindrical coordinates
\begin{equation}
 r^lY_{lm}(\theta,\phi) = (-1)^m \sqrt{\frac{(2l+1)(l+m)!(l-m)!}{4\pi}} e^{im\phi} \sum_k \frac{(-1)^k {r_\mathrm{c}}^{2k+m}  z^{l-2k-m}}{2^{2k+m}(m+k)!k!(l-m-2k)!} \, , \label{CylindricalSolidHarmonic}
\end{equation}
where $k$ is summed over all values where each factorial is non-negative. Here we are careful with our notation such that $r$ and $r_\mathrm{c}$ are the radial distances in the spherical and cylindrical coordinate systems respectively, $\phi$ is the azimuthal angle for both coordinate systems,  $\theta$ is the spherical polar angle, and $z$ is vertical position.

For simplicity, we will calculate all closed forms for $m\geq0$. The inner multipole moments for negative $m$ can easily be calculated with the following identity
\begin{equation}
 q_{l(-m)} = (-1)^mq^\ast_{lm}\,.
\end{equation}

\subsection{Inner multipoles of a cylinder}

\begin{figure}[t]
\centering
\includegraphics[width=.9\textwidth]{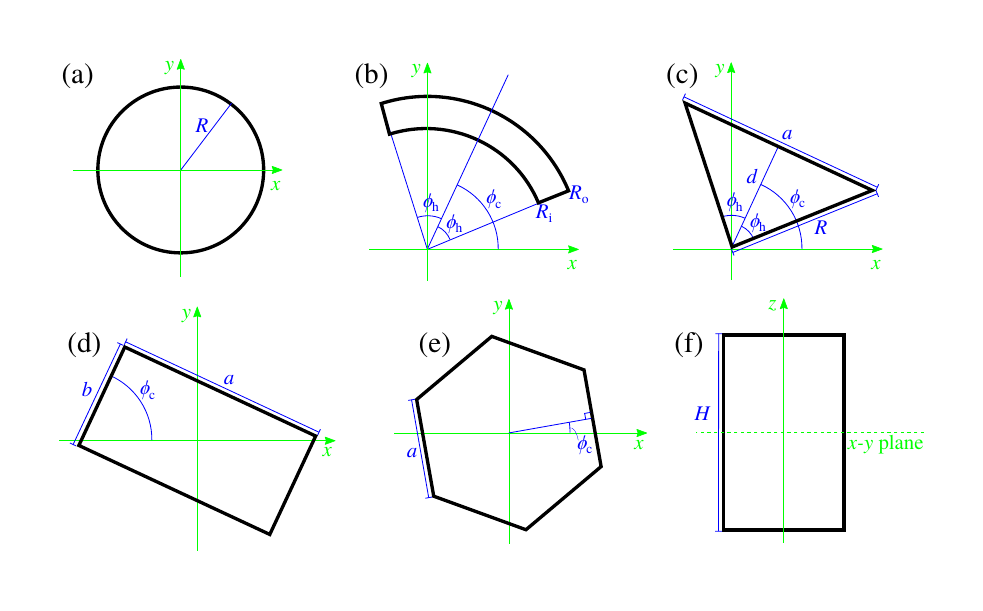}
\caption{(a)--(e) Cross sections of cylinder, annular section, isosceles triangular prism, cuboid, and $N$-sided regular polygonal prism respectively. (f) Side view for all aforementioned prisms.}
\label{FIG:Shapes}
\end{figure}

From symmetry we can see that the inner multipoles $q_{lm}$ of a homogeneous cylinder of density $\rho$ requires $m=0$ due to rotational azimuthal symmetry and for $l$ to be even from vertical symmetry. Using Eqn.~\ref{CylindricalSolidHarmonic} and integrating over the volume of the cylinder with radius $R$ and height $H$ centred on the origin (See Figure~\ref{FIG:Shapes}(a) and (f))
\begin{align}
 q_{l0} = \rho\int_{V_c} r^lY^{\ast}_{l0}(\theta,\phi)\cald{V_c} &=  \rho\sqrt{\frac{2l+1}{4\pi}}l! \sum_{k=0}^{l/2} \frac{(-1)^k}{2^{2k}k!k!(l-2k)!} \int\displaylimits^{H/2}_{-H/2}\int\displaylimits^R_0 \int\displaylimits^{2\pi}_0  {r_\mathrm{c}}^{2k} z^{l-2k} r_\mathrm{c}\cald{\phi}\cald{r_\mathrm{c}}\cald{z} \nonumber\\
 &= M\sqrt{\frac{2l+1}{4\pi}} \frac{l!}{2^{l}} \sum_{k=0}^{l/2} \frac{(-1)^k   R^{2k} H^{l-2k}  }{k!(k+1)!(l-2k+1)!}\,,\\&\qquad\qquad\qquad\qquad\text{for $l$ even,}\nonumber
 \label{CylindricalInnerMultipole}
\end{align}
where $M$ is the mass of the cylinder. This result is consistent with that derived by Lockerbie, Veryaskin, and Xu  \cite{Lockerbie1993} and has the useful form of being the mass of the object multiplied by a geometrical factor. For efficient programming this equation can easily be written as a simple recursion relation:
\begin{align}
 q_{l0}  &= M\sqrt{\frac{2l+1}{4\pi}} \sum_{k=0}^{l/2} S(l,k)\,,
\end{align}
where
\begin{align}
 S(0,0) &= 1\,,\\ S(l+2,0) &= \frac{(l+1)H^2}{4(l+3)}S(l,0)\,,\\ S(l,k+1) &= -\frac{(l-2k+1)(l-2k)}{(k+1)(k+2)}\frac{R^2}{H^2}S(l,k)\,.
 \label{CylindricalRecursion}
\end{align}

\subsection{Inner multipoles of an annular section}\label{AnnularSec}

A more generalised case for the cylinder is an annular section with inner radius $R_\mathrm{i}$ and outer radius $R_\mathrm{o}$ which extends over the azimuthal angular range from $\phi_\mathrm{c}-\phi_\mathrm{h}$ to $\phi_\mathrm{c}+\phi_\mathrm{h}$ (See Figure~\ref{FIG:Shapes}(b) and (f)). The $z$ integral can be solved separately (\ref{App:zInt}). The integral to solve is then
\begin{equation}
 \int\displaylimits^{R_\mathrm{o}}_{R_\mathrm{i}} \int\displaylimits^{\phi_\mathrm{c}+\phi_\mathrm{h}}_{\phi_\mathrm{c}-\phi_\mathrm{h}} e^{-im\phi} {r_\mathrm{c}}^{2k+m}  r_\mathrm{c}\cald{\phi}\cald{r_\mathrm{c}} = 
 \frac{2\left({R_\mathrm{o}}^{2k+m+2} - {R_\mathrm{i}}^{2k+m+2}\right)}{2k+m+2}
 \begin{cases}
e^{-im\phi_\mathrm{c}}\frac{\sin(m\phi_\mathrm{h})}{m} &\text{for } m\neq0\\
\phi_\mathrm{h}&\text{for } m=0
\end{cases}
  \,.
\end{equation}
From~\ref{App:zInt} we know that from vertical symmetry that $(l-m)$ must be even. We then combine the above result with Eqn.~\ref{zPrism}, and the other terms for $q_{lm}$ in front of the integral.  To write the multipole as the mass multiplied by a geometric factor we need to factor out the volume $\phi_\mathrm{h}({R_\mathrm{o}}^{2} - {R_\mathrm{i}}^{2})H$,
\begin{align}
 q_{lm}= M &\sqrt{\frac{(2l+1)(l+m)!(l-m)!}{4\pi}}  e^{-im\phi_\mathrm{c}}\sinc(m\phi_\mathrm{h}) \times \nonumber\\ &\sum_{k=0}^{(l-m)/2}  \frac{ (-1)^{k+m} H^{l-2k-m} }{2^{l-1}  k!(m+k)!(l-m-2k+1)!(2k+m+2) }\times\nonumber\\ &\left(\frac{{R_\mathrm{o}}^{2k+m+2} - {R_\mathrm{i}}^{2k+m+2}}{{R_\mathrm{o}}^{2} - {R_\mathrm{i}}^{2}}\right)\,, \qquad\text{for $(l-m)$ even, and $m\geq0$}\,,
\end{align}
where we note that using the sinc function removes the need for separate cases for $m=0$ and $m\neq0$. This equation can be shown to be consistent with the results given in Adelberger {\it et al.}  \cite{Adelberger2006}.

\subsection{Inner multipoles of an isosceles triangular prism}
Here we define an isosceles triangle using the same geometry as the annular section except with only one radius $R$, with $\phi_\mathrm{h}< \tfrac{\pi}{2}$ (See Figure~\ref{FIG:Shapes}(c) and (f)). Using the solution for the $z$-integral for a prism (\ref{App:zInt}), the remaining integrals to solve are
\begin{equation}
 \int\displaylimits^{\frac{R\cos\phi_\mathrm{h}}{\cos(\phi-\phi_\mathrm{c})}}_{0} \int\displaylimits^{\phi_\mathrm{c}+\phi_\mathrm{h}}_{\phi_\mathrm{c}-\phi_\mathrm{h}} e^{-im\phi} {r_\mathrm{c}}^{2k+m}  r_\mathrm{c}\cald{\phi}\cald{r_\mathrm{c}} = 
 \frac{(R\cos\phi_\mathrm{h})^{2k+m+2}}{2k+m+2}\int\displaylimits^{\phi_\mathrm{c}+\phi_\mathrm{h}}_{\phi_\mathrm{c}-\phi_\mathrm{h}}
 \frac{e^{-im\phi}}{\cos^{2k+m+2}(\phi-\phi_\mathrm{c})} \cald{\phi},
\end{equation}
which is solved in~\ref{App:IntFlat}. Factoring out the mass $M = \rho HR^2\cos^2\phi_\mathrm{h}\tan\phi_\mathrm{h}$,
\begin{align}
q_{lm}= M&\sqrt{\frac{(2l+1)(l+m)!(l-m)!}{4\pi}}\frac{e^{-im\phi_\mathrm{c}}}{2^{l-1}}\sum_{k=0}^{(l-m)/2} \frac{(-1)^{k+m}{H}^{l-2k-m} (R\cos\phi_\mathrm{h})^{2k+m}}{(m+k)!k!(l-m-2k+1)!(2k+m+2)}\times \nonumber\\&
\sum_{p=0}^{\left \lfloor{m/2}\right \rfloor } (-1)^p
\binom{m}{2p}\sum_{j=0}^k \binom{k}{j}\frac{\tan^{2j+2p}\phi_\mathrm{h}}{2j+2p+1}\,, \qquad\text{for $(l-m)$ even, and $m\geq0$}\,,
\end{align}
where $\left \lfloor{\tfrac{m}{2}}\right \rfloor$ denotes rounding $\tfrac{m}{2}$ down to the nearest integer. Calling the base of the triangle $a=2R\sin\phi_\mathrm{h}$ and the shortest line to the base $d=R\cos\phi_\mathrm{h}$, a more simple form is:
\begin{align}
q_{lm}= M&\sqrt{\frac{(2l+1)(l+m)!(l-m)!}{4\pi}} \frac{e^{-im\phi_\mathrm{c}}}{2^{l-1}}
\sum_{k=0}^{(l-m)/2} \frac{(-1)^{k+m}{H}^{l-2k-m} d^{2k+m}}{(m+k)!k!(l-m-2k+1)!(2k+m+2)}\times \nonumber\\&
\sum_{p=0}^{\left \lfloor{m/2}\right \rfloor } (-1)^p
\binom{m}{2p}\sum_{j=0}^k \binom{k}{j}\frac{1}{2j+2p+1}\left(\frac{a}{2d}\right)^{2j+2p}\,, \qquad\text{for $(l-m)$ even, and $m\geq0$}\,.
\end{align}

\subsection{Inner multipoles of a cuboid}

A cuboid can be described as a sum of two pairs of isosceles triangular prisms. Defining a cuboid of height ($z$-axis) $H$ to be consistent with the above prisms, the other other two sides $a$ and $b$ are defined such that when $\phi_\mathrm{c}=0$, $a$ is parallel to the $y$-axis and $b$ is parallel to the $x$-axis (See Figure~\ref{FIG:Shapes}(d) and (f)). By symmetry we can see that the each pair of isosceles triangles are offset by an angle $\pi$ therefore $m$ must always be even for a nonzero multipole. As with all prisms centred in $z$, $(l-m)$ must be even, and therefore $l$ is also even. The inner multipoles for a cuboid are thus:
\begin{align}
q_{lm}= M&\sqrt{\frac{(2l+1)(l+m)!(l-m)!}{4\pi}} (-1)^{m/2}  e^{-im\phi_\mathrm{c}}
\times\nonumber\\&\sum_{k=0}^{(l-m)/2} \frac{(-1)^{k}{H}^{l-2k-m} }{(m+k)!k!2^{l+2k+m}(l-m-2k+1)!(2k+m+2)} \times\nonumber\\
&\sum_{p=0}^{m/2 } (-1)^p
\binom{m}{2p}\sum_{j=0}^k \binom{k}{j}\frac{
a^{2k+m-2j-2p}b^{2j+2p}+b^{2k+m-2j-2p}a^{2j+2p}
}{2j+2p+1}\,,\nonumber\\&
 \qquad\qquad\qquad\qquad\qquad\qquad\text{for both $m$ and $l$ even, and $m\geq0$}\,.
\end{align}

\subsection{Inner multipoles of an $N$-sided regular polygonal prism}

Consider an $N$-sided regular polygonal prism, with height $H$ with its centre of figure at the origin. The angle between the right-most side (in the $xy$-plane) and the $y$-axis is $\phi_\mathrm{c}$. The side length of the polygon is $a$ (See Figure~\ref{FIG:Shapes}(e) and (f)). The inner multipole moments can easily be calculated by combining the results for $N$ identical isosceles triangular prisms each rotated by an angle $\frac{2\pi}{N}$ with respect to the last. By symmetry, the angular term for the $N$ prisms add to $N$ if $m$ is a multiple of $N$, or else it vanishes, hence the moment is simply (again non-zero for $(l-m)$ is even):
\begin{align}
q_{lm}= M&\sqrt{\frac{(2l+1)(l+m)!(l-m)!}{4\pi}} e^{-im\phi_\mathrm{c}} \times \nonumber\\
&\sum_{k=0}^{(l-m)/2} \frac{(-1)^{k+m}{H}^{l-2k-m} a^{2k+m}}{(m+k)!k!2^{l+2k+m-1}(l-m-2k+1)!(2k+m+2)} \sum_{p=0}^{\left \lfloor{m/2}\right \rfloor } (-1)^p \binom{m}{2p}\times \nonumber\\
&\sum_{j=0}^k \binom{k}{j}\frac{\tan^{2j+2p-2k-m}\left(\frac{\pi}{N}\right)}{2j+2p+1}\qquad\qquad\text{for $(l-m)$ even, and $m = 0,N,2N,\ldots$}\,.
\end{align}

\subsection{Inner multipoles of an azimuthal section of a cone}

\begin{figure}[t]
\centering
\includegraphics[width=.6\textwidth]{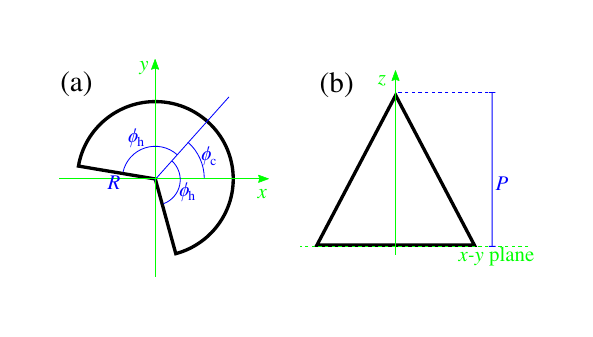}
\caption{(a) Base of azimuthal section of a cone. (b) Side view of cone.}
\label{FIG:Cone}
\end{figure}

Consider a cone with a base centred at the origin with a radius $R$, the apex of the cone is on the $z$-axis with $z=P$. The cone is defined in the azimuthal angular range from $\phi_\mathrm{c}-\phi_\mathrm{h}$ to $\phi_\mathrm{c}+\phi_\mathrm{h}$ (See Figure~\ref{FIG:Cone}). The azimuthal integral for the inner multipoles was already solved in Section~\ref{AnnularSec}. The radial and $z$ integrals are:
\begin{align}
\int\displaylimits_0^P \int\displaylimits_0^{R-\frac{zR}{P}} {r_\mathrm{c}}^{2k+m+1}z^{l-2k-m} \cald{r_\mathrm{c}}\cald{z} &= \frac{(R/P)^{2k+m+2}}{2k+m+2}\int\displaylimits_0^P (P-z)^{2k+m+2}z^{l-2k-m}\cald{z}\\
&= \frac{(2k+m+1)!(l-2k-m)!}{(l+3)!} R^{2k+l+2}P^{l-2k-m+1}\,.
\end{align}
We can therefore write the inner multipole moments as
\begin{align}
q_{lm}= 3M & \sqrt{\frac{(2l+1)(l+m)!(l-m)!}{4\pi}} (l+3)! e^{-im\phi_\mathrm{c}}\sinc(m\phi_\mathrm{h})\times\nonumber\\ &\sum_{k=0}^{\lfloor(l-m)/2\rfloor} \frac{(-1)^{k+m} (2k+m+1)!R^{2k+m}P^{l-2k-m}}{2^{2k+m-1}(m+k)!k!}\,, \qquad\text{for $m\geq0$}\,.
\end{align}

\section{Inner multipoles of an inhomogeneous cylinder}

A number of precision measurements of the universal constant of gravitation use cylindrical masses as the primary source and test masses. Density gradients in these masses have been measured and approximations have been used to calculate the effect on measurements\cite{Quinn2014,Parks2014}. Here we present a closed form expression for a cylindircal mass with a linear density gradient. The density can be written as
\begin{equation}
 \rho(\vec{r}) = \rho_0 + \rho_\mathrm{r}r_\mathrm{c}\sin(\phi+\phi_\mathrm{I}) + \rho_\mathrm{z}z
\end{equation}
where $\phi_\mathrm{I}$ is the direction of the radial gradient. Multipole from this cylinder can be divided into three calculations, the first being the homogeneous cylinder, the remaining two being the effect of the radial and vertical gradients described below.

\subsection{Radial gradient}

For a cylinder of height $H$ and radius $R$ centred on the origin the radial inner multipole moment from inhomogeneous densities is calculated as:
\begin{align}
 q_{lm} = (-1)^m \rho_\mathrm{r}\sqrt{\frac{(2l+1)(l+m)!(l-m)!}{4\pi}}  \sum_k \frac{(-1)^k }{2^{2k+m}(m+k)!k!(l-m-2k)!}\times\nonumber\\
 \int\displaylimits^{H/2}_{-H/2}\int\displaylimits^R_0\int\displaylimits^{2\pi}_0 e^{-im\phi}\sin(\phi+\phi_\mathrm{I}){r_\mathrm{c}}^{2k+m+2}  z^{l-2k-m} \cald{\phi}\cald{r_\mathrm{c}}\cald{z}
\end{align}
The radial portion of this integrates to zero for $m\neq\pm1$. For for $m=1$ the integral is trivial:
\begin{align}
 q_{l1} = i\pi\rho_\mathrm{r} e^{i\phi_\mathrm{I}} \sqrt{\frac{(2l+1)(l+1)!(l-1)!}{4\pi}}  \sum_{k=0}^{(l-1)/2} \frac{(-1)^{k} R^{2k+4}H^{l-2k} }{2^{l}(k+1)!k!(l-2k)!(2k+4)}\\
 \text{For $l$ odd, otherwise 0}\nonumber
\end{align}

\subsection{Vertical gradient}
In the case of the vertical gradient the integral is also trivial, and by symmetry only $m=0$ terms are nonzero
\begin{align}
 q_{l0} = \rho_\mathrm{z}  \sqrt{\frac{(2l+1)}{4\pi}}\;l!  \sum_{k=0}^{l/2}  \frac{(-1)^k R^{2k+2}  H^{l-2k+2}}{2^{l+1} (k!)^2(l-2k)!(2k+2)(l-2k+2)}\\
 \text{For $l$ odd, otherwise 0}\nonumber
\end{align}

\section{Discussion}

Care must be taken, however, when performing numerical calculations. First many programming languages define $\sinc(x)$ as $\frac{\sin(\pi x)}{\pi x}$ rather than $\frac{\sin(x)}{x}$. Also, for large degree multipole moments, numerical rounding errors become significant as the sum over $k$ has terms with alternating sign which individually can be many orders of magnitude larger than the final result. As a rule of thumb we find that for $l\gtrsim50$ quadruple-precision floats should be used for calculations requiring precision better than 1 part in $10^6$. Using quadruple-precision floating point operations we have found results are still accurate beyond double-precision for $l>100$. This can be checked on an individual basis by comparing the ratio of the magnitude of largest term in the sum over $k$ and the sum itself to the numerical precision of the data type used. For example, we can write an inner multipole moment of an object as
\begin{equation}
 q_{lm} = A(\cdots) \sum_k (-1)^k B_k(\cdots)\,,
\end{equation}
where $A$ and $B$ are functions of the variables needed to describe the object. We can then estimate the relative error in our numerical calculations as
\begin{equation}
 \frac{\Delta q_{lm}}{q_{lm}} \sim \frac{\max({B_k}(\cdots))}{ \sum_k (-1)^k B_k(\cdots)} P_{B_k},
\end{equation}
where $P_{B_k}$ is the numerical precision of the floating point data type used to store $B_k$ and its sum ($P_{B_k} \sim 10^{-16}$ for double-precision and $P_{B_k} \sim 10^{-34}$ for quadruple-precision). This estimate assumes all other sources of numerical error are negligible.

\section{Conclusion}
\begin{figure}[t]
\centering
\includegraphics[width=.4\textwidth]{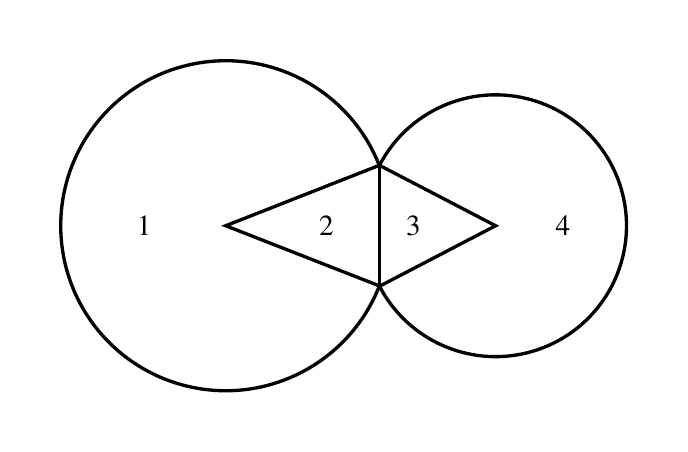}
\caption{Top view of two overlapping holes which can be modelled as two cylindrical sections in the angular range without overlap, plus two isosceles triangular prisms.}
\label{FIG:Overlap}
\end{figure}
We have derived close form expressions for the gravitational inner multipole moments for a number of homogeneous elementary solids in terms of their mass multiplied by a geometrical factor. We have also derived expressions for the gravitational inner multipole moments of a cylinder with a linear density gradient. Using the translation and rotation equations for multipoles, a number of more complex but commonly occurring shapes can be modelled. For example, overlapping cylindrical holes can be modelled as two cylindrical sections in the angular range without overlap, plus two isosceles triangular prisms, all with negative mass (see Figure~\ref{FIG:Overlap}); any irregular polygon prism can be modelled as a combination of isosceles triangular prisms; or a truncated cone can be modelled as one cone subtracted from another. The equations provided are relatively simple to code to allow multipole calculations of Newtonian gravitational interactions between complex structures to any desired degree.

\appendix
\section{Integrals}

\subsection{$z$-Integral for prisms}\label{App:zInt}

For prismatic solids, the $z$-integral can be solved separately from the other two coordinates:
\begin{align}
 f_{lmk}(H):=\frac{1}{2^{2k+m}(l-m-2k)!}\int\displaylimits^{H/2}_{-H/2} z^{l-2k-m} \cald{z} &=  \frac{\left[\left(\frac{H}{2}\right)^{l-2k-m+1}-\left(\frac{-H}{2}\right)^{l-2k-m+1}\right]}{2^{2k+m}(l-m-2k)!(l-2k-m+1)}\,.
 \end{align}
The integral vanishes if $(l-m)$ is odd, therefore:
\begin{align}
f_{lmk}(H)&=   \frac{{H}^{l-2k-m+1}}{2^{l}(l-m-2k+1)!}\,,\label{zPrism}\\&\qquad\qquad\text{for $(l+m)$  even, and zero otherwise.}\nonumber 
\end{align}

\subsection{Integral used for flat sides}\label{App:IntFlat}

To integrate the radial coordinate over a flat edge the following integral must be solved:
\begin{equation}
g_{km}(\phi_\mathrm{c},\phi_\mathrm{h}):=\int\displaylimits^{\phi_\mathrm{c}+\phi_\mathrm{h}}_{\phi_\mathrm{c}-\phi_\mathrm{h}}
\frac{\displaystyle e^{-im\phi}}{\cos^{2k+m+2}(\phi-\phi_\mathrm{c})}
\cald{\phi}\,.
\end{equation}
Substituting $\phi_i = \phi-\phi_\mathrm{c}$ for symmetry, then
\begin{equation}
e^{-im\phi_\mathrm{c}}\int\displaylimits^{\phi_\mathrm{h}}_{-\phi_\mathrm{h}}
\frac{\displaystyle e^{-im\phi_i}}{\cos^{2k+m+2}\phi_i}
\cald{\phi_i} =
e^{-im\phi_\mathrm{c}}\int\displaylimits^{\phi_\mathrm{h}}_{-\phi_\mathrm{h}}
\frac{\displaystyle \cos(m\phi_i)}{\cos^{2k+m+2}\phi_i}
\cald{\phi_i},\label{DiffIntSub1}
\end{equation}
where the imaginary part of the integral is odd and therefore evaluates to zero.

But for $m\geq0$
\begin{equation}
\cos(m\phi_i) = \sum_{p=0}^{\left \lfloor{m/2}\right \rfloor } (-1)^p
\binom{m}{2p}
\cos^{m-2p}\phi_i\sin^{2p}\phi_i\,.
\end{equation}
Substituting this into Eqn.~\ref{DiffIntSub1} gives
\begin{equation}
g_{km}(\phi_\mathrm{c},\phi_\mathrm{h}) = e^{-im\phi_\mathrm{c}}\sum_{p=0}^{\left \lfloor{m/2}\right \rfloor } (-1)^p
\binom{m}{2p}
\int\displaylimits^{\phi_\mathrm{h}}_{-\phi_\mathrm{h}}
\frac{\displaystyle \tan^{2p}\phi_i}{(\cos^2\phi_i)^{k+1}}
\cald{\phi_i}\,.\label{DiffIntSum}
\end{equation}
The integral in the sum can be rewritten as
\begin{equation}
\int\displaylimits^{\phi_\mathrm{h}}_{-\phi_\mathrm{h}}
\frac{\displaystyle (1+\tan^2\phi_i)^k\tan^{2p}\phi_i}{\cos^2\phi_i}
\cald{\phi_i},
\end{equation}
using the identity $1+\tan^2\phi_i = \frac{1}{\cos^2\phi_i}$. If we substitute $x=\tan\phi_i$, then
\begin{equation}
\int\displaylimits^{\tan\phi_\mathrm{h}}_{-\tan\phi_\mathrm{h}}(1+x^2)^kx^{2p}
\cald{x} = 2\sum_{j=0}^k \binom{k}{j}\frac{\tan^{2j+2p+1}\phi_\mathrm{h}}{2j+2p+1}\,,\label{IntFromSum}
\end{equation}
and therefore by substituting Eqn.~\ref{IntFromSum}  into Eqn.~\ref{DiffIntSum}, we conclude that
\begin{align}
\int\displaylimits^{\phi_\mathrm{c}+\phi_\mathrm{h}}_{\phi_\mathrm{c}-\phi_\mathrm{h}}
\frac{\displaystyle e^{-im\phi}}{\cos^{2k+m+2}(\phi-\phi_\mathrm{c})}
\cald{\phi} = 2e^{-im\phi_\mathrm{c}}\sum_{p=0}^{\left \lfloor{m/2}\right \rfloor } (-1)^p
&\binom{m}{2p}\sum_{j=0}^k \binom{k}{j}\frac{\tan^{2j+2p+1}\phi_\mathrm{h}}{2j+2p+1}\,,\nonumber\\
& \text{for } m\geq0\,.
\end{align}

\end{document}